\documentclass[a4paper,conference]{IEEEtran}
\ifCLASSINFOpdf
\else
\fi
\usepackage{cite}
\usepackage[hyphens]{url} 
\usepackage{graphicx} 

\usepackage{caption} 
\usepackage{subcaption}

\usepackage{algorithm}
\usepackage{algpseudocode}

\usepackage{comment}

\usepackage{xcolor}
\usepackage{amsfonts}
\usepackage{array}
\usepackage{amsmath}

\def\BibTeX{{\rm B\kern-.05em{\sc i\kern-.025em b}\kern-.08em
    T\kern-.1667em\lower.7ex\hbox{E}\kern-.125emX}}
\newcolumntype{C}[1]{>{\centering\let\newline\\\arraybackslash\hspace{0pt}}p{#1}}

\usepackage{amssymb,amsmath}

\begin{document}
%
\title{Learning to Adapt to Domain Shifts with Few-shot Samples in Anomalous Sound Detection}

\author{\IEEEauthorblockN{Bingqing Chen\IEEEauthorrefmark{2}\IEEEauthorrefmark{4},
Luca Bondi\IEEEauthorrefmark{4}, 
and Samarjit Das\IEEEauthorrefmark{4}} 
\IEEEauthorblockA{\IEEEauthorrefmark{2} 
Carnegie Mellon University, Pittsburgh, PA, USA; \IEEEauthorrefmark{4} Bosch Research \& Technology Center, Pittsburgh, PA, USA}
\texttt{bingqinc@andrew.cmu.edu, \{luca.bondi, samarjit.das\}@us.bosch.com}}


%


\maketitle

\begin{abstract}
Anomaly detection has many important applications, such as monitoring industrial equipment. Despite recent advances in anomaly detection with deep-learning methods, it is unclear how existing solutions would perform under out-of-distribution scenarios, e.g., due to shifts in machine load or environmental noise. 
Grounded in the application of machine health monitoring, we propose a framework that adapts to new conditions with few-shot samples. 
Building upon prior work, we adopt a classification-based approach for anomaly detection and show its equivalence to mixture density estimation of the normal samples. We incorporate an episodic training procedure to match the few-shot setting during inference. We define multiple auxiliary classification tasks based on meta-information and leverage gradient-based meta-learning to improve generalization to different shifts. We evaluate our proposed method on a recently-released dataset of audio measurements from different machine types. It improved upon two baselines by around 10\% and is on par with best-performing model reported on the dataset.  
\end{abstract}


%
\IEEEpeerreviewmaketitle

\section{Introduction}
\label{sec:intro}

Anomaly detection is the task of identifying anomalous observations~\cite{ivanovska2020,wang2018,prado2016,feng2010,reif2008}. In this paper, we focus on anomaly detection applied to machine health monitoring via audio signal. Detecting anomalies is useful for identifying incipient machine faults, condition-based maintenance, and quality assurance, which are integral components towards Industry 4.0. In comparison to direct measurements, audio is a cost-effective, non-intrusive, and scalable sensor modality. 

\begin{figure}
{ 
\includegraphics[width= 1.\linewidth]{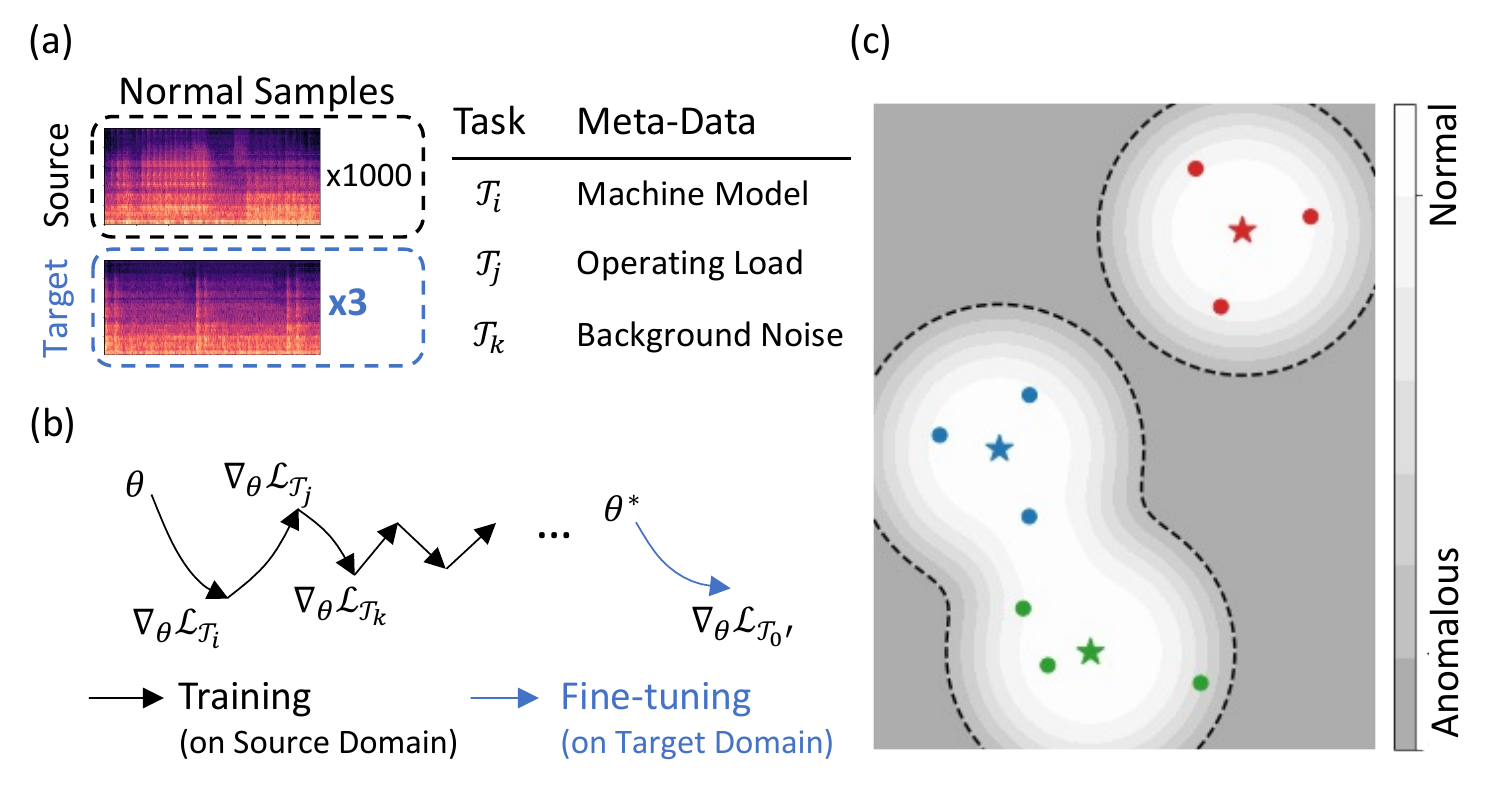}%
}
\caption{Framework. (a) Our anomalous sound detection system adapts to new conditions with 3-shot samples. We adopt a classification-based approach and define the auxiliary classification tasks on available meta-data. (b) In the outer loop, we alternative between the auxiliary classification tasks, such that the model can quickly adapt to new conditions.
(c) For a given auxiliary classification task, the anomaly score is calculated based on distance between each sample ($\cdot$) to the prototypes ($\star$) on the embedding space. This is equivalent to mixture density estimation on normal samples. }
\label{fig:framework}
\vspace{-0.5cm} 
\end{figure}

Specifically, we focus on the problem set-up (Figure \ref{fig:framework}a) where our system adapts to new conditions using only a handful of samples (specifically 3 in the experiment). It is clearly not desirable for an observation to be tagged as anomalous due to changes in operating condition or environmental noise. Thus, we aim to develop an anomalous sound detection (ASD) system that can adapt to new conditions quickly with few-shot samples. Additionally, some meta-information may be available, e.g., machine model and operating load.

Due to the practical difficulty of enumerating potential anomalous conditions and generating such samples, it is typically assumed that only normal samples are available for training \cite{ruff2018deep}. This poses the challenge that anomaly detectors need to learn to identify anomalies in the absence of any anomalous samples. Furthermore, deep anomaly detectors perform in unexpected and not well-understood ways under out-of-distribution scenarios  \cite{ruff2021unifying}. Last but not least, the unique characteristics of audio add to the challenge. 

Given its superior performance, especially on similar problems \cite{koizumi2020description, giri2020self, morita2021anomalous}, we adopt a classification-based approach for anomaly detection \cite{bergman2020classification}, where auxiliary classification tasks are defined on the available meta-data. However, existing classification-based methods \cite{bergman2020classification, kawaguchi2021description} are not designed for anomaly detection under domain shift, and thus are not sufficient for competitive performance on their own, which we show in Section \ref{sec:results}. Thus, we propose a novel approach, with bi-level optimization, to tackle the challenging problem of anomaly detection under domain shifts.  In the outer loop (Figure \ref{fig:framework}b), we alternate between all available auxiliary classification tasks leveraging a gradient-based meta-learning (GBML) algorithm \cite{nichol2018first}, such that the resulting model can quickly adapt to the target domain. In the inner loop (Figure \ref{fig:framework}c), we train the classification-based anomaly detector  with an episodic procedure to match the few-shots setting \cite{snell2017prototypical}.
While classification-based methods show strong empirical results in the literature \cite{golan2018deep, hendrycks2018deep, bergman2020classification}, there is not a convincing explanation for that performance. Thus, we propose a supplementary explanation that classification-based methods are equivalent to mixture density estimation on normal samples. 
Finally, we evaluate our proposed approach on a recently-released dataset of audio measurements from a variety of machine types \cite{kawaguchi2021description} and show strong empirical results.

\section{Related Work}
\label{sec:review}

\paragraph{Anomaly Detection}
Recently, there is a surge of interest in using deep learning approaches for anomaly detection to handle complex, high-dimensional data. Anomaly detection methods can be categorized as density estimation-based, classification-based, and reconstruction-based \cite{bergman2020classification,ruff2021unifying}. Density estimation-based methods fit a probabilistic model on normal data, and data from low density region are considered anomalies. For the purpose of anomaly detection, we are only interested in level sets of the distribution, and thus classification-based methods learn decision boundaries that delineate high-density and low-density region. Reconstruction-based methods train a model to reconstruct normal samples, and use reconstruction error as a proxy for anomaly score. 


Due to the superior performance of classification-based methods on our application of interest \cite{giri2020self, morita2021anomalous, koizumi2020description}, we focus our review on this class of methods and refer interested readers to comprehensive reviews \cite{chalapathy2019deep, ruff2021unifying} for more information. A representative method is Deep Support Vector Data Description (SVDD) \cite{ruff2018deep}, a one-class classification method, where the neural network learns a representation that enforces the majority of normal samples to fall within a hypersphere. The core challenge of this line of work is learning the decision boundary for a binary classification problem in the presence of only normal samples. Alternatively, outlier exposure  \cite{hendrycks2018deep} takes  advantage of an auxiliary dataset of outliers, i.e. samples disjoint from both normal and anomalous ones, to improve performance for anomaly detection and generalization at unseen scenarios. Finally, other methods train models on auxiliary classification tasks and use the negative log-likelihood of a sample belonging to the correct class as the anomaly score. Some examples of such auxiliary classification tasks include identifying the geometric transformation applied to the original sample \cite{golan2018deep, bergman2020classification} and classifying the metadata associated with the sample \cite{giri2020self, morita2021anomalous}. This is reminiscent of self-supervised learning (SSL), where learning to distinguish between self and others is conducive to learning salient representation for other downstream tasks.

\paragraph{Meta-Learning}
The objective of meta-learning is to find a model that can generalize to a new, unseen task with a handful of samples \cite{finn2017model}. 
Metric-based meta-learning algorithms learn  feature representation such that query samples can be classified based on their similarity to the support samples with known labels. For instance, Matching Networks \cite{vinyals2016matching} use attention mechanism to evaluate the similarity. Prototypical Networks \cite{snell2017prototypical} establish class prototypes based on support samples, and assign each query sample to its nearest prototype. 

GBML algorithms, popularized by  model-agnostic meta-learning (MAML) \cite{finn2017model}, train models that can quickly adapt to new tasks with gradient-based updates, typically using a bi-level optimization procedure. Variants of MAML have been proposed in works such as \cite{nichol2018first, raghu2019rapid, wang2021bridging}.

\paragraph{Anomalous Sound Detection}
In comparison to general anomaly detection, audio has its unique characteristics, e.g., temporal structure  \cite{rushe2019anomaly}. On the same task, \cite{lopez2021ensemble} observed that there was limited performance gain by trying a number of common domain adaptation techniques, highlighting the challenge and the need for research on audio-specific methods.  

While reconstruction-based \cite{rushe2019anomaly, suefusa2020anomalous} and density estimation-based \cite{yamaguchi2019adaflow} anomaly detection methods have been used for anomalous sound detection, classification-based methods  enjoy superior performance on similar / the same tasks \cite{koizumi2020description, giri2020self, morita2021anomalous}. Albeit differences in specific implementation, the core idea is to classify metadata, such as machine identification, and compute the anomaly score from the negative log-likelihood of class assignment. Also relevant to the task are Sniper \cite{koizumi2019sniper} and SpiderNet \cite{koizumi2020spidernet}, which memorize few-shot anomalous samples to prevent overlooking other anomalous samples. In particular, SpiderNet uses attention mechanism to evaluate the similarity with known anomalous samples, similar to Matching Networks.

\section{Approach}
\label{sec:approach}

\subsection{Preliminaries}
By defining the distribution of normal data as $\mathbb{P}^\star$ over the data space $\mathcal{X}\subseteq\mathbb{R}^D$, anomalies may be characterized as the set where normal samples is unlikely to be, i.e. $\mathcal{A} = \{x\in \mathcal{X}\;|\;\mathbb{P}^\star(x)\leq \alpha\}$, where $\alpha$ is a threshold \cite{ruff2021unifying} that controls Type I error. 

A fundamental assumption in anomaly detection is \textit{the concentration assumption}, i.e. the region where normal data reside can be bounded \cite{ruff2021unifying}. More precisely, there exists $\alpha \geq 0$, such that 
$\mathcal{X} \setminus \mathcal{A} = \{x\in \mathcal{X}\;|\;\mathbb{P}^\star(x) > \alpha  \}$ is nonempty and small. 
Note that this assumption does not require the support of $\mathbb{P}^\star$ be bounded; only that the support of the high-density region of $\mathbb{P}^\star$ be bounded. In contrast, anomalies need not be concentrated and a commonly-used assumption is that anomalies follow a uniform distribution over  $\mathcal{X}$ \cite{ruff2021unifying}. 

\subsection{Problem Formulation}

We study the problem of anomaly detection under domain shift, having access to only few-shot samples in the target domain. 
Specifically, we have access to a dataset from the source domain $\mathcal{D} = \{(x_i, y_i)\}$, where each $x_i\in \mathbb{R}^D$ is a normal sample and $y_i\in \mathbb{R}^L$  are the labels for auxiliary classification tasks.  We also have access to a small number of samples in the target domain $\mathcal{D}' = \{(x'_i, y_i)\}$, where $x'_i\in \mathbb{R}^D$ is a normal sample from a domain-shifted condition. Note that we only access to normal samples in both source and target domain. We denote each auxiliary task as $\mathcal{T}_l$ and the set of $L$ auxiliary tasks as $\mathcal{T}= \{\mathcal{T}_1, \dots, \mathcal{T}_L\}$. 

We use a neural network $f_\theta:\mathbb{R}^{D}\rightarrow \mathbb{R}^d$ to map samples from the data space to an embedding space, where $D\gg d$. We denote the embedding that corresponds to each sample as $z_i$, where $z_i =f_\theta(x_i) \in \mathbb{R}^d$, and the distribution of normal data in the embedding space as $\mathbb{P}^\star_z$.

\subsection{Anomaly Detection via Metadata Classification}

To simplify notation, we consider a single auxiliary task $\mathcal{T}_l$ with class labels $1, \dots, K$ in this subsection. We denote $\mathcal{D}_k=\{(x_i, y_i)\in \mathcal{D}\;|\;y_i=k)\}$. 

As discussed in Section \ref{sec:review},  anomaly detection methods based on auxiliary classification train models to differentiate between classes, either defined via metadata inherent to the dataset or synthesized by applying various transformations to the sample, and calculate the anomaly score as the negative log-likelihood of a sample belonging to the correct class (Eqn. \ref{eq:anomaly_score}) \cite{hendrycks2016baseline}.
\begin{equation}\label{eq:anomaly_score}
    \Omega_\theta(x_i, y_i) =-\log \mathbb{P}_\theta(y=y_i|x_i)  
\end{equation}

Following \cite{bergman2020classification}, we define the likelihood by distance on the embedding space (Eqn. \ref{eq:prob}), rather than a classifier head, to handle new, unseen classes in the target domain, where $d$ is a distance metric and $c_k= \frac{1}{|\mathcal{D}_k|}\sum_{(x_i, y_i)\sim \mathcal{D}_k} f_\theta(x_i)$ is the class centroid. 
\begin{equation}\label{eq:prob}
\mathbb{P}_\theta(y=k|x_i)  = \frac{\exp(-d(f_\theta(x_i), c_k))}{\sum_{k'}\exp(-d(f_\theta(x), c_{k'}))}
\end{equation}
The learning objective is to maximize the log-likelihood of assigning each sample to its correct class, or equivalently minimizing  the negative log-likelihood as in Eqn. \ref{eq:ce_loss}, leading to the familiar cross-entropy loss.  
\begin{equation}\label{eq:ce_loss}
\mathcal{L}_{\tau_l}(x_i, y_i; \theta) 
= -\log\frac{\exp(-d(f_\theta(x_i), c_{y_i}))}{\sum_{k}\exp(-d(f_\theta(x), c_{k}))}
\end{equation}

An explanation for the strong empirical performance of anomaly detection based on auxiliary classification is that the auxiliary classification tasks are conducive to learning salient feature useful for anomaly detection \cite{golan2018deep, giri2020self}. However, this explanation provides limited insight in how to define the auxiliary classification tasks and what kind of  performance may be  expected. 

Here, we provide an alternative explanation that the classification objective (Eqn. \ref{eq:ce_loss}) is equivalent to mixture density estimation on normal samples. In Deep SVDD \cite{ruff2018deep}, it is assume that the neural network can find a latent representation such that the majority of normal points fall within a single hypersphere or equivalently $\mathbb{P}_z^\star$ follows an isotropic Gaussian distribution. Based on the same intuition, we 
assume $\mathbb{P}_z^\star$ can be characterized by a mixture model (Equation \ref{eq:mixture_model}), where $\pi$ is the prior distribution for class membership. As an example, a machine may operate normally under different operating loads. Instead of assuming that normal data can be modeled by a single cluster, the distribution of normal data  can be better characterized with a number of clusters each corresponding to an operating load.  We hypothesize that this more flexible representation enables the model to learn fine-granular features conducive to anomaly detection, and  extrapolate better to unseen scenarios.
\begin{equation}\label{eq:mixture_model}
\hat{\mathbb{P}}^\star_z (z)  = \sum_{k=1}^K \pi_k \mathbb{P}(z|y=k)
\end{equation}
We also assume that $\mathbb{P}(z|y=k)$ can be characterized by a distribution in the exponential family, $\mathbb{P}(z|y=k)  \propto \exp(-d(z, c_k)) $
where $d$ is a Bregman divergence \cite{banerjee2005clustering}. The choice of $d$ dictates the modeling assumption on the conditional distribution, $\mathbb{P}(z|y)$. For instance, by choosing squared Euclidean distance, i.e. $d(z, c_k) = ||z-c_k||^2_2$, one models each cluster as an isotropic Gaussian. Given these modeling assumptions, it is easy to see that  
\begin{equation}\label{eq:cluster_assignment}
\mathbb{P}_\theta(y=k|x_i)  = \frac{\pi_k\exp(-d(f_\theta(x_i), c_k))}{\sum_{k'}\pi_{k'}\exp(-d(f_\theta(x), c_{k'}))}
\end{equation}

Eqn. \ref{eq:cluster_assignment} is the same as Eqn. \ref{eq:prob}, by assuming a flat prior on class membership, which can be satisfied by sampling mini-batches with balanced classes. This shows that learning the auxiliary classification task is equivalent to performing mixture density estimation with exponential family, where each cluster corresponds to a class.




\begin{algorithm}[tb]
\caption{Learning to Adapt with Few-shot Samples}
\label{alg:algorithm}
\begin{algorithmic}[1]
\State \textbf{input}: Data $\mathcal{D}$, $\mathcal{D}'$; Test set $\mathcal{D}_{\text{test}}$; Model $f_\theta$;
\State \textbf{parameter}:  Learning rate $\alpha$;  Outer step-size $\epsilon$; Number of inner and fine-tuning iterations $T$ and $T_{\text{test}}$ 

\Function{ComputeLoss}{$\mathcal{S}$, $\mathcal{Q}$, $l$} 
    \State//  \emph{input: support set, query set, task index}
    \State //  \emph{Compute prototypes from the support set}
    \State  $c_k= \frac{1}{|\mathcal{S}_k|}\sum_{(x_i, y_i)\sim \mathcal{S}_k}f_\theta (x_i), \; \forall k$     
    \State //  \emph{Calculate task loss from the query set}
\State \Return  $\frac{1}{|\mathcal{Q}|}\sum_{(x_i, y_i)\sim \mathcal{Q}} \mathcal{L}_{\tau_l}(x_i, y_i; \theta)$
\EndFunction
\State 
\Procedure{Train}{$f_\theta$, $\mathcal{D}$} 
    \State//  \emph{input: model, training data}
  \State \textbf{initialize} $\theta$
    \While{not done}
        \For{$\mathcal{T}_l\sim \mathcal{T}$} 
        \State \textbf{set} $\theta^{(0)} = \theta$
        \For{$t = 0, \dots, T-1$}
        \State // \emph{Sample support and query set}
        \State $\mathcal{S}, \mathcal{Q}\sim \mathcal{D}$
        \State $\mathcal{L}_{\mathcal{T}_l}$ = \Call{ComputeLoss}{$\mathcal{S}$, $\mathcal{Q}$, $l$}
        \State \textbf{update} $\theta^{(t+1)} = \theta^{(t)} - \alpha \nabla_\theta \mathcal{L}_{\mathcal{T}_l}(\theta^{(t)})$
        \EndFor
        \State \textbf{meta update} $\theta \leftarrow\theta + \epsilon\left[\theta^{(T)}-\theta\right]$
        \EndFor
    \EndWhile
\EndProcedure
\State
\Procedure{Inference}{$f_{\theta^\star}$, $\mathcal{D}'$, $\mathcal{D}_{\text{test}}$}
\State //  \emph{input: trained model, few-shot examples, test set}
\State \textbf{set} $\theta^{(0)} = \theta^\star$
\State //  \emph{Fine-tuning on few-shot examples}
\For{$t = 0, \dots, T_{\text{test}}-1$}
        \State $\mathcal{L}_{\mathcal{T}_0}$ = \Call{ComputeLoss}{$\mathcal{D}'$, $\mathcal{D}'$, 0}
        \State \textbf{update} $\theta^{(t+1)} = \theta^{(t)} - \alpha \nabla_\theta \mathcal{L}_{\mathcal{T}_0}(\theta^{(t)})$
\EndFor
\State //  \emph{Compute anomaly score for test samples}
\State \textbf{compute} $\Omega_{\theta^{(T_{\text{test}})}}(x_i, y_i), \;\; \forall (x_i, y_i)\sim \mathcal{D}_{\text{test}}$
\EndProcedure 
\end{algorithmic}
\end{algorithm}
\paragraph{Adaptation with Few-shot Samples} \label{sec:Protonet} A distinction from the typical anomaly detection problem set-up is that we need to adapt to new conditions and only have access to few-shot samples from the target domain. Thus, we draw from the few-shot classification literature, and modify the classification-based anomaly detector for the few-shot setting. Specifically, Prototypical networks (ProtoNet) \cite{snell2017prototypical} is a few-shot learning method that also classifies samples based on distance on the embedding space. During inference, ProtoNet 
uses the few-shot examples to establish the new class centroid (i.e. prototype) under domain-shifted conditions, $c'_k$, i.e. 
$c'_k = \frac{1}{|\mathcal{D}'_k|}\sum_{(x'_i, y'_i)\sim \mathcal{D}'_k}f_\theta (x'_i)$. Each test sample is assigned to the nearest class prototype, with the same probability as defined by Eqn. \ref{eq:prob}.

ProtoNet adopts an episodic training procedure (see \textsc{ComputeLoss} in Algorithm \ref{alg:algorithm}), such that the training condition matches the test condition. During training, ProtoNet splits each mini-batch into support set, $\mathcal{S}$, and query set, $\mathcal{Q}$, where the support set simulates the few-shot examples, and the query set simulates the test samples during inference. In other words, the prototypes are computed on the support set and the loss is evaluated on the query set.

\paragraph{Outlier Exposure} We also use the outlier exposure (OE) technique \cite{hendrycks2018deep} to boost performance. It is commonly assumed that anomalies follow a uniform distribution over the data space \cite{ruff2021unifying}. Thus, the auxiliary loss for outlier exposure $\mathcal{L}_{\text{OE}}$ is defined as cross-entropy to the uniform distribution, and added to the learning objective with weight $\lambda$. 
$$\mathcal{L} = \mathcal{L}_{\mathcal{T}_l} + \lambda\mathcal{L}_{\text{OE}}$$

\subsection{Multi-objective Meta-learning} 
So far, we have focused the discussion on the case of there being a single auxiliary classification task. But, more meta-information regarding operating conditions may be available. Since the samples may be subject to different changes due to operating condition, machine load, or environment noise,  we hypothesize that training on a variety of tasks is conducive to generalizing well to different domain shifts. Also, it is not known a priori which auxiliary classification task would be most effective for anomaly detection. Thus, it is sensible to train on all available auxiliary classification tasks. Empirically, we show in Section \ref{sec:results} that training on all auxiliary classification tasks does outperform training on any single one. 

Recall that meta-learning trains the model on a distribution of tasks such that it can quickly learn  new, unseen tasks with few-shot samples. Thus, these auxiliary classification tasks can be naturally incorporated into meta-learning algorithms. Wang et al. in \cite{wang2021bridging} draw the close connections between multi-task learning and meta-learning. 
A distinction, however, is that meta-learning typically trains on different tasks of the same nature, e.g. 5-way image classification, while multi-task learning may train on functionally related tasks of different nature, e.g. image reconstruction and classification. Our approach falls under the latter case. 

We use Reptile \cite{nichol2018first}, a first-order variant of MAML, which learns a parameter initialization that can be fine-tuned quickly on a new task. Reptile repeatedly samples a task $\mathcal{T}_l\sim \mathcal{T}$, trains on it, and moves the model parameter towards the trained weights on that task (see \textsc{Train} in Algorithm \ref{alg:algorithm}) following
$$\theta \leftarrow\theta + \epsilon\left[\theta^{(T)}-\theta\right]$$
where $\theta^{(T)}$ are the model parameters after training on $\mathcal{T}_l$ for $T$ gradient steps and $\epsilon$ is the step-size for meta-update. Based on Taylor series analysis,  \cite{nichol2018first} shows that Reptile simultaneously minimizes expected loss over all tasks, and maximizes within-task generalization, i.e. taking a gradient step for a specific task on one mini-batch, also improves performance on other tasks.  

 During inference, the trained model is first fine-tuned on few-shot examples from the target domain, and the anomaly score (Eqn. \ref{eq:anomaly_score}) can be computed on the test set (see \textsc{Inference} in Algorithm \ref{alg:algorithm}).   

\begin{table*}[ht]
    \centering
    \begin{tabular}{c C{5.0cm} C{5.5cm} C{3.8cm}}
         \hline \noalign{\smallskip}
         \textbf{Machine Type} & \textbf{Anomalous Conditions} & \textbf{Variations across Domains} & \textbf{Auxiliary Classification Tasks} \\\hline \noalign{\smallskip}
         \textbf{Toy Car}& Bent shaft; Deformed / melted gears;  Damaged wheels &  Car model; Speed;  Microphone type and position & Section ID; Model No.; Speed  \\ \noalign{\smallskip}
         \textbf{Toy Train} & Flat tire; Broken shaft; Disjoint railway track; Obstruction on track & Train model; Speed; Microphone type and position & Section ID; Model No.; Speed \\ \noalign{\smallskip} 
         \textbf{Fan} & Damaged / unbalanced wing; Clogging; Over-voltage& Wind strength; Size of the fan; Background noise & Section ID \\\noalign{\smallskip} 
         \textbf{Gearbox} & Damaged gear; Overload; Over-voltage & Voltage; Arm-length; Weight& Section ID; Voltage; Arm length; Weight; \\ \noalign{\smallskip}
         \textbf{Pump} & Contamination; Clogging; Leakage; \newline Dry run & Fluid viscosity; Sound-to-Noise Ratio (SNR); Number of pumps & Section ID \\ \noalign{\smallskip}
         \textbf{Slider} &Damaged rail; Loose belt; No grease & Velocity; Operation pattern; Belt material & Section ID; Slider type; Velocity; Displacement; \\ \noalign{\smallskip}   
         \textbf{Valve} &Contamination&Operation pattern; Existence of pump in background; Number of valves & Section ID; Pattern; Existence of pump; Multiple valves\\ \hline
    \end{tabular}
    \caption{Summary of anomalous conditions, domain shifts, and auxiliary classification tasks}
    \label{tab:dataset}
    \vspace{-0.5cm} 
\end{table*}

\section{Experiment}
\label{sec:experiments}
In this section, we describe the experimental set-up, including the dataset, evaluation metrics, our implementation details, baselines, and ablation study. 
\paragraph{Dataset}
We use the dataset from Detection and Classification of Acoustic Scenes and Events (DCASE) Challenge 2021 Task 2 \cite{kawaguchi2021description}, which is composed of subsets of ToyADMOS2 \cite{harada2021toyadmos2} and MIMII DUE \cite{tanabe2021mimii}. The dataset consists of normal and anomalous samples from seven distinct machine types, i.e. toy car, toy train, fan, gearbox, pump, slider, and valve. Anecdotally, we found it extremely challenging to distinguish normal and anomalous with untrained human ears. The anomalous samples are generated by intentionally damaging the machines in different ways (see Table \ref{tab:dataset}). Note that the anomalous samples are used for evaluation only. 

Each sample is a 10s audio clip at 16kHz sampling rate, including both machine sound and environment noise. 
For each machine type, the data is grouped into 6 sections, where a \texttt{section} is a unit for performance evaluation and roughly corresponds to a distinct machine. In each section, the samples are collected from two different conditions, which we refer to as the source and target domain. The domain shifts are different across sections.  A notable challenge is that there are only 3 normal samples for the target domain in each section, while there are 1000  for the source domain. Table \ref{tab:dataset} summarizes the domain shifts in the dataset.

For evaluation, each section has 100 normal samples and 100 anomalous samples for both source and target domain. Section 0, 1, 2 are designated as the validation set, and Section 3, 4, 5 are designated as the the test set. 

\paragraph{Evaluation Metrics}   We follow the same evaluation procedure as the DCASE Challenge, and report the Area Under the Receiver Operating Characteristic curve (AUROC) and the partial AUROC (pAUROC), which controls the false positive rate at 0.1. The metrics are aggregated over sections and machines types with harmonic mean. We focus on the results on the target domain.

\paragraph{Implementation Details}
Our preprocessing and model architecture follows the 2\textsuperscript{nd} baseline from the challenge organizer \cite{kawaguchi2021description}. The raw audio is preprocessed into log-mel-spectrogram with a frame size of 64ms, a hop size of 50\%, and 128 mel filters. After preprocessing, 64 consecutive frames in a context window are treated as a sample, and samples are generated from a audio clip by shifting the context window by 8 frames. Each 10s audio clip results in 32 samples, with each sample, $x_i\in \mathbb{R}^{64\times 128}$. We use MobileNetV2 \cite{sandler2018mobilenetv2} as backbone, and set the bottleneck size to 128. Thus, $z_i = f_\theta(x_i) \in \mathbb{R}^{128}$. 

We follow the optimization procedure in \cite{nichol2018first}. In the inner loop, we use ADAM as the optimizer with a learning rate $\alpha = 0.001$, $\beta_1 = 0$, and $\beta_2 = 0.999$. Each mini-batch is sampled to be have balanced class, with 3 audio clips as support and 5 as query for each class, simulating the 3-shot setting during inference. For the outer loop, we use SGD with step-size, $\epsilon$, linearly annealing from 1 to 0. The number of iterations for inner loop and fine-tuning are $T = 8$, and $T_{\text{test}} = 50$ respectively. We have a model for each machine type and train it for 10K steps. For outlier exposure, we generate outliers by 1) taking samples from machines other than the one being trained, and 2) synthesizing samples via frequency warping following \cite{giri2020self}. We use $\lambda=0.2$. 

The implementation is in \texttt{PyTorch}\cite{pytorch} 
and trained on a machine with Intel\textsuperscript{\textregistered} Core\texttrademark~i9-10900KF CPU@3.70GHz and NVIDIA GeForce RTX\texttrademark~3090 GPU.

\paragraph{Baselines and Ablation Study} We compare our model to the two baselines provided by the challenge organizer \cite{kawaguchi2021description}. The 1\textsuperscript{st} baseline adopts a reconstruction-based approach with an autoencoder. The 2\textsuperscript{nd} baseline uses a classification-based approach with MobileNetV2 as the backbone. The auxiliary classification task is defined as section ID. Note that our preprocessing and the neural architecture is the same as that of the 2\textsuperscript{nd} baseline. We also compare our model to the best-performing system \cite{lopez2021ensemble} among the 77 submissions to the challenge, which used an ensemble of two classification-based models and one density estimation-based model. We conduct ablation study to analyze the contribution of individual components.

\section{Results}\label{sec:results}

\begin{figure}
    \centering
    \includegraphics[width = \linewidth]{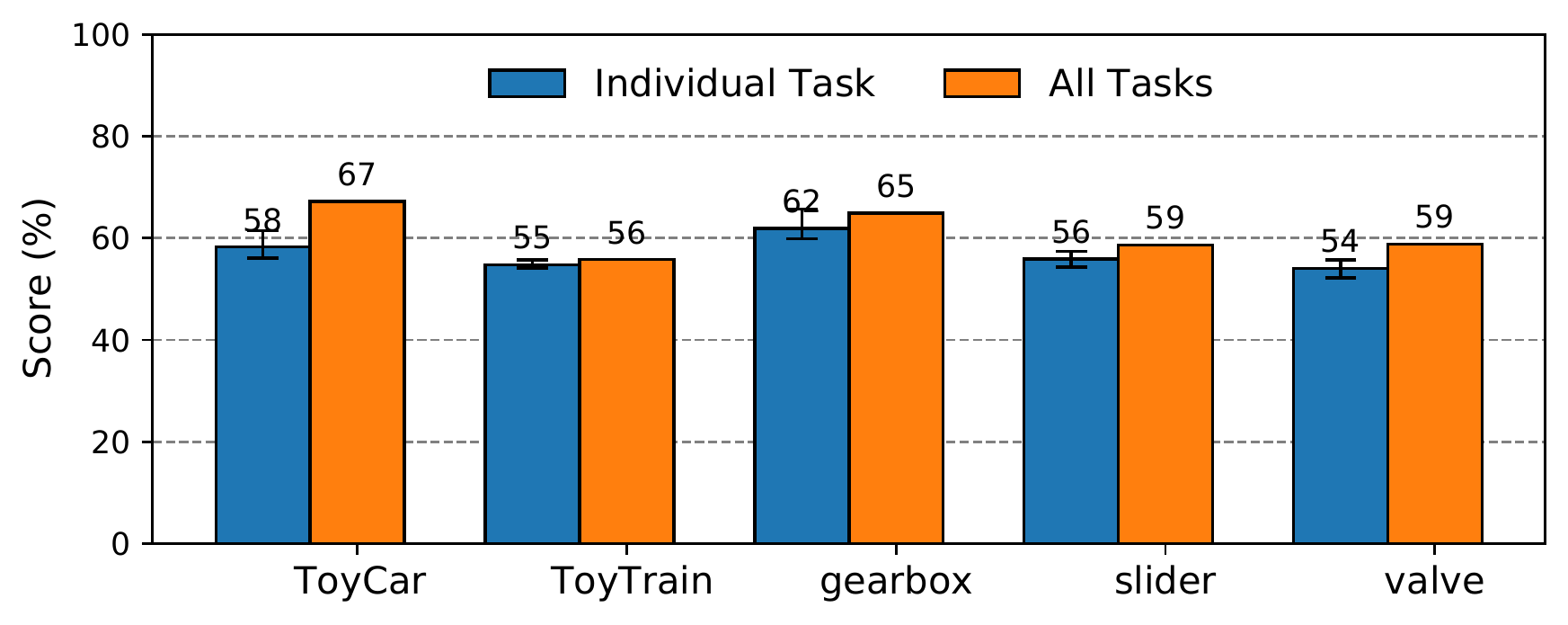}
    \caption{Performance comparison of using  individual vs. all auxiliary classification tasks (evaluated on validation set). For individual tasks, the bar length indicates the averaged score over tasks and error bar indicates the minimum and the maximum over tasks.}
    \label{fig:by_task}
    \vspace{-0.5cm}
\end{figure}

\paragraph{Choice of Auxiliary Classification Tasks}
While in prior work \cite{giri2020self, morita2021anomalous, lopez2021ensemble}, it is popular to use machine / section ID as the auxiliary classification task, we hypothesize that training on multiple auxiliary classification tasks performs better than training on any individual one. We define auxiliary classification tasks by parsing the  meta-information associated with each audio clip, also summarized in Table \ref{tab:dataset}. Take \texttt{Toy Car} as an example, we have access to information on car model and speed. Intuitively, being able to differentiate between car models/speeds,  is conducive for the model to adapt to new model/speed and potentially other new conditions. 
\begin{figure*}[ht]
    \centering
    \includegraphics[width=\linewidth]{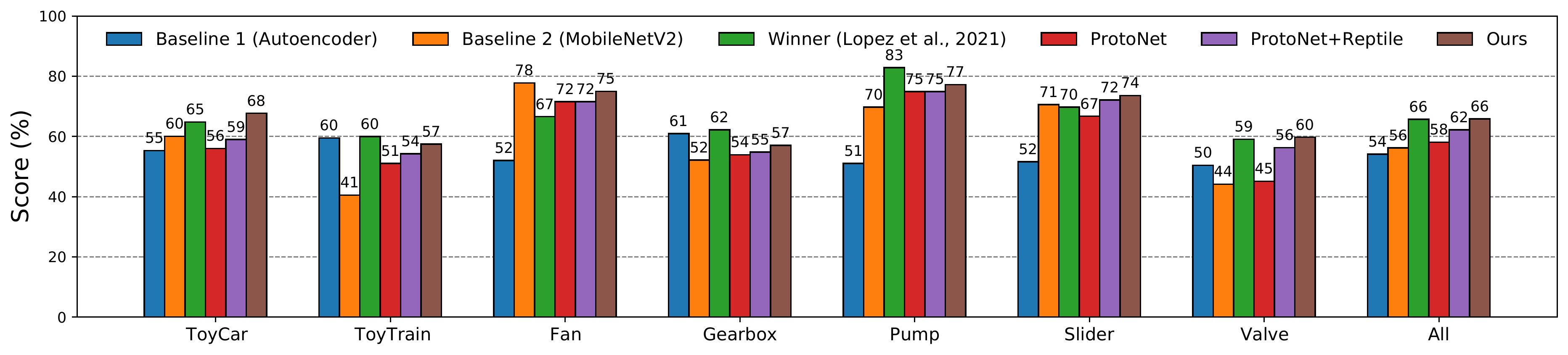}
    \caption{Performance comparison with baselines and ablation (evaluated over the test set). The proposed method (Ours) is compared against Baseline 1 and 2,  DCASE 2021 Task2 winner, and ProtoNet, and our proposed method without outlier exposure (ProtoNet + Reptile).}
    \label{fig:results}
    \vspace{-0.5cm} 
\end{figure*}

To confirm the hypothesis, we train the anomaly detector on individual classification tasks with ProtoNet, and compare its performance with our proposed approach of alternating between all auxiliary classification tasks using Reptile. We report the score on the validation set in Figure \ref{fig:by_task}. As mentioned in Section \ref{sec:experiments}, we follow the evaluation procedure in the DCASE challenge and report the performance as harmonic mean over AUROC and pAUROC@0.1 of relevant data sections.
\texttt{Fan} and \texttt{Pump} are not compared here, as only one auxiliary classification task is available for these two machine types. As expected, training on all tasks performs consistently better across different machine types, confirming the hypothesis.


\paragraph{Overall Results} 

The overall results evaluated on the test set are summarized in Figure \ref{fig:results}. In summary, our model (\texttt{Ours}) improved on average upon Baseline 1 and 2 by 11.7\% and 9.6\% respectively, and is on par with the best-performing system \cite{lopez2021ensemble}, despite being a third of its model complexity (measured by the number of model parameters). While Baseline 1 and Baseline 2 perform similarly based on the harmonically-averaged scores, there are significant variations across machines. It appears reconstruction-based Baseline 1 and classification-based Baseline 2 excel on different machines. Our proposed model outperforms both baselines for \texttt{ToyCar}, \texttt{Pump}, \texttt{Slider}, and \texttt{Valve}.

In comparison to vanilla classification-based approach in Baseline 2, we trained the classifier episodically to match few-shot setting (\texttt{ProtoNet}), and iterated among different auxiliary classification tasks (\texttt{ProtoNet + Reptile}) to improve generalization. Finally, we augmented the dataset via OE, adding up to our proposed approach (\texttt{Ours}). On average, the episodic training procedure improved performance by 1.9\%, GBML applied to auxiliary classification improved performance by 4.1\% and data augmentation by OE improved performance by 3.6\%. The most significant improvement comes from training on all auxiliary classification tasks via Reptile. There is no improvement from \texttt{Fan} or \texttt{Pump} as these two machine types have only one auxiliary classification task.

\paragraph{Low-Dimensional Visualization} Qualitatively, we show the embedding of the test sample from the trained model for \texttt{Toy Car}, before and after fine-tuning. The stars indicate the prototypes established by the 3-shot samples. 
\begin{figure}
    \centering
    \subcaptionbox{Before Fine-tuning \label{fig:Before}}{\includegraphics[width = \linewidth]{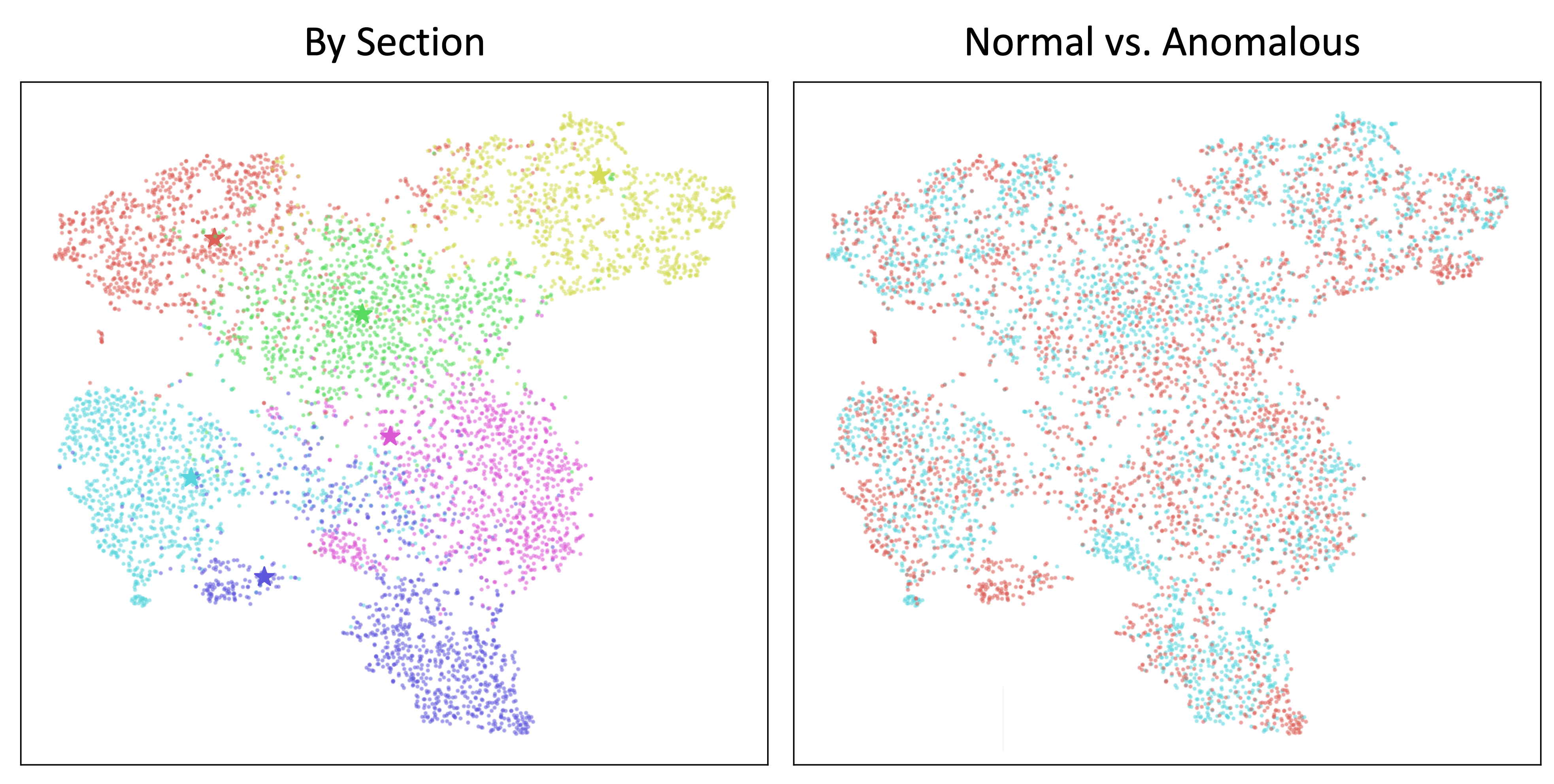}}
    \subcaptionbox{After Fine-tuning \label{fig:After}}{\includegraphics[width=\linewidth]{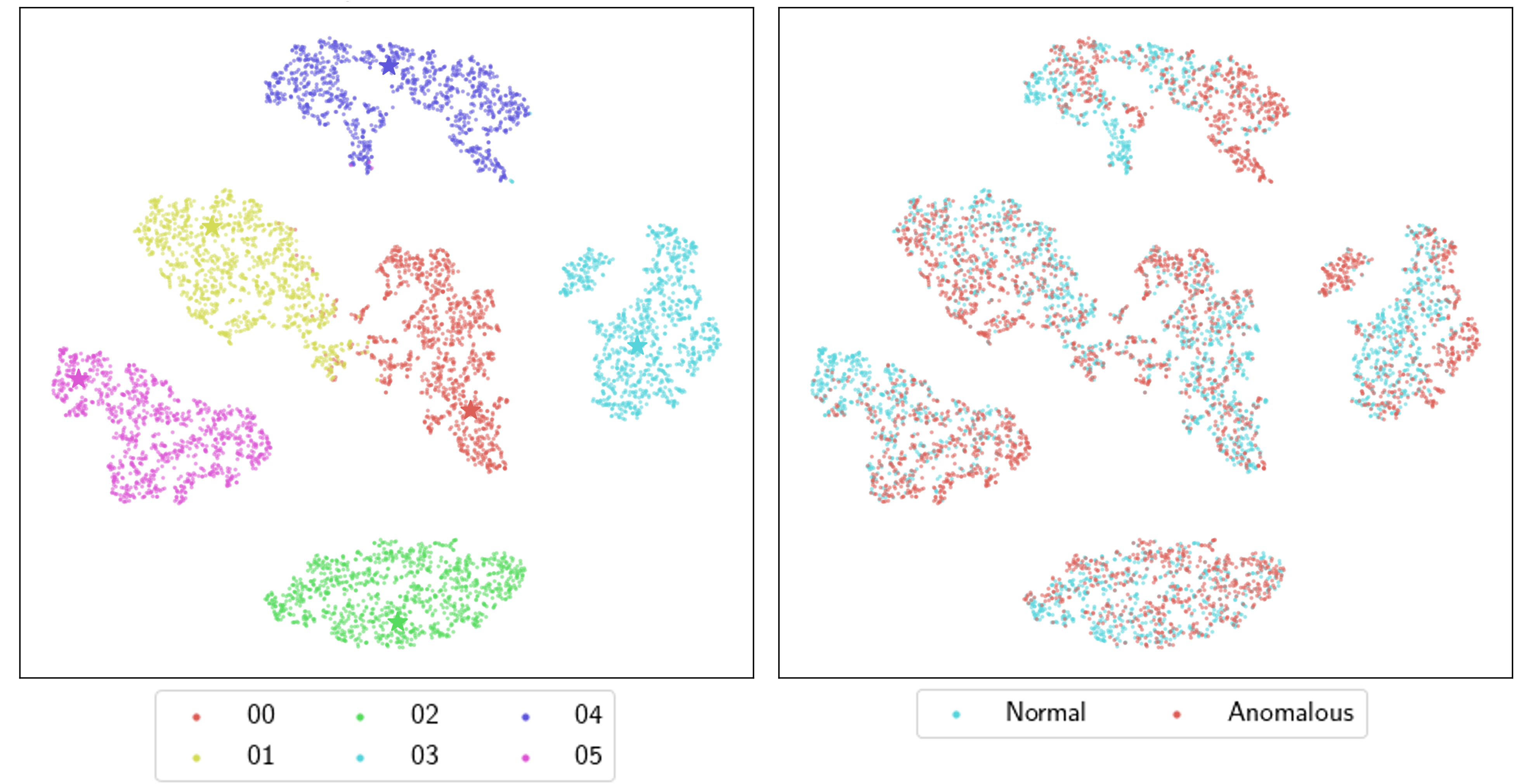}}
    \caption{Trained Embedding of Toy Car (via t-SNE) on the test set before \textit{(top)} and after \textit{(bottom)} fine-tuning on the target domain $\mathcal{D'}$.  Fine-tuning improves discriminability of normal vs. anomalous samples \textit{(bottom-right)} on the target domain.}
   \vspace{-0.5cm} 
    \label{fig:my_label}
\end{figure}

The trained model has not seen any samples from the target domain prior to fine-tuning. Regardless, the model has already learned meaningful embedding (Figure \ref{fig:Before}), where the samples are naturally separated by section. But, the normal and anomalous samples are not yet  well separated. During fine-tuning, the prototypes attract normal samples, and as a result the normal and anomalous samples become better separated. 

\section{Conclusions}
In this work, we tackle the challenging task of adapting unsupervised anomaly detector to new conditions using few-shot samples. To achieve this objective, we leverage approaches from meta-learning literature. We train a  classification-based anomaly detector in an episodic procedure to match the few-shot setting during inference. We use a GBML algorithm to find parameter initialization that can quickly adapt to new conditions with gradient-based updates. Finally, we boost our model with outlier exposure.

Grounded in the application of machine health monitoring, we evaluate our proposed method on a recently-released dataset of audio measurements from different machine types, used for DCASE challenge 2021. We model is on par with best-performing system among the 77 submissions to the challenge. We conduct ablation to analyze the contribution of each component, and the GBML procedure leads to the most significant improvement.


\bibliographystyle{IEEEtran}
\bibliography{ref} 


\end{document}